\documentclass{emulateapj}
\def\fdg{\hbox{$.\!\!^\circ$}}

\def\farcm{\hbox{$.\mkern-4mu^\prime$}}

\def\bik{\mbox{\boldmath $k$}}

\def\bip{\mbox{\boldmath $p$}}
\def\bir{\mbox{\boldmath $r$}}
\def\biv{\mbox{\boldmath $v$}}

\def\bix{\mbox{\boldmath $x$}}

\def\MW{\mathcal{W}}

\usepackage{float}
\usepackage{amsmath}
\usepackage{epsfig,floatflt}
\usepackage{subfigure}

\received{July 12, 2005}
\accepted{September 18, 2005}
\begin{document}

\title{Power Spectrum of Cosmic Momentum Field  
       Measured from the SFI Galaxy Sample}
\author{Chan-Gyung Park and Changbom Park}\affil{Korea Institute for Advanced Study, 130-722, Seoul, Korea}
\email{parkc@kias.re.kr;cbp@kias.re.kr}

%\author{Chan-Gyung Park}
%\affil{KIAS,Korea} 
%\email{parkc@kias.re.kr}

%\date{Received March 1, 2004 / Accepted May 20, 2004}

\begin{abstract}
We have measured the cosmic momentum power spectrum from the peculiar
velocities of galaxies in the SFI sample. The SFI catalog contains field spiral
galaxies with radial peculiar velocities derived from the $I$-band
Tully-Fisher relation.
As a natural measure of the large-scale peculiar velocity field,
we use the cosmic momentum field
that is defined as the peculiar velocity field weighted by local number
of galaxies.
We have shown that the momentum power spectrum can be derived from
the density power spectrum for the constant linear biasing of galaxy
formation, which makes it possible to estimate
$\beta_{\rm S} = \Omega_m^{0.6} / b_{\rm S}$ parameter precisely
where $\Omega_m$ is the matter density parameter and $b_{\rm S}$
is the bias factor for optical spiral galaxies.
At each wavenumber $k$ we estimate $\beta_{\rm S}(k)$ as the ratio
of the measured to the derived momentum power over a wide range of scales
($0.026~h^{-1}\textrm{Mpc} \lesssim k \lesssim 0.157~h^{-1}\textrm{Mpc}$)
that spans the linear to the quasi-linear regimes.
The estimated $\beta_{\rm S}(k)$'s have stable values around $0.5$,
which demonstrates the constancy of $\beta_{\rm S}$ parameter at scales
down to $40~h^{-1}\textrm{Mpc}$.
We have obtained $\beta_{\rm S}=0.49_{-0.05}^{+0.08}$ or
$\Omega_m = 0.30_{-0.05}^{+0.09} b_{\rm S}^{5/3}$, and
the amplitude of mass fluctuation as
$\sigma_8\Omega_m^{0.6}=0.56_{-0.21}^{+0.27}$.
The 68\% confidence limits include the cosmic variance.
We have also estimated the mass density power spectrum. For example,
at $k=0.1047~h\textrm{Mpc}^{-1}$ ($\lambda=60~h^{-1}\textrm{Mpc}$)
we measure $\Omega_m^{1.2}P_{\delta}(k)=(2.51_{-0.94}^{+0.91})
\times 10^3$ $(h^{-1}\textrm{Mpc})^3$, which is lower compared to
the high-amplitude power spectra found from the previous maximum likelihood
analyses of peculiar velocity samples like Mark III, SFI, and ENEAR.
\end{abstract}


\section{Introduction}
\label{sec_introduction}

Peculiar motions of galaxies in the nearby Universe are very powerful tool
for examining the underlying mass fluctuations on large scales.
The galaxy peculiar velocity directly probes the large-scale matter density
field, and thus gives biasing information of galaxy distribution.
The advantage in using the peculiar velocity data is that we can explore
the large-scale structures in the real space rather than in the redshift space,
without redshift space distortion.
On the other hand, the galaxy peculiar velocities or the absolute distances
have large errors proportional to distance because the absolute
distance is usually inaccurately measured compared with the redshift.
For this reason peculiar velocity samples are in general much smaller
than redshift survey samples. In spite of this disadvantage,
the peculiar velocity sample is very useful since the observed velocity field
contains information on larger scales compared with the density field
in a given survey volume.

In the study of peculiar velocity field, homogeneous peculiar
velocity sample with well-defined selection criteria is essential
because the sparse and inhomogeneous sample significantly induces biases
in estimating physical parameters.
Over a decade or so there has been major progress in the peculiar velocity
studies. Observationally, there has been a dramatic improvement
with the completion of large Tully-Fisher (TF) and $D_{\rm n}$--$\sigma$
redshift-distance surveys in both hemispheres.
Those are the Mark III catalog \citep{wil97a} that contains about 3,300
spiral and elliptical galaxies, the SFI catalog \citep{gio94} 
with about 1300 late-type spiral galaxies with $I$-band TF distance estimates,
and the ENEAR catalog \citep{dac00b} containing about 1400 early-type
galaxies with $D_{\rm n}$--$\sigma$ distances.
Recently, new distance indicators such as the Type Ia supernovae
\citep{rie97,rad04} and the surface brightness fluctuation of the early-type
galaxies \citep{bla99} have been used in the peculiar velocity study.

Theoretically, there have been many works on the peculiar velocity fields
(\citealt{str95}; \citealt{zar02a} for a review). 
The galaxy peculiar velocities have been used to reconstruct
the three-dimensional velocity field or the real-space matter density field
of the local Universe (e.g., \citealt{dac96,dek99}), and to estimate
the amplitude of mass fluctuation and/or $\beta=\Omega_m^{0.6}/b$ parameter,
where $b$ is the linear bias factor.
Among the many analysis methods that have been developed and applied to
the observational data, one of the major development has been the POTENT method
\citep{ber90,dek99} which reconstructs the smoothed three-dimensional
peculiar velocity field from the observed radial velocity data.  
As a new method of estimating $\beta$ parameter accurately, 
\citet[hereafter P00]{par00} has introduced the cosmic momentum field
and developed the momentum power spectrum analysis method
which does not require a smoothing of galaxy peculiar velocities.
Other studies have applied the velocity correlation function statistic
\citep{gor89,bor00a,bor00b}, the maximum likelihood (ML) method
\citep{fre99,sil01}, the orthogonal mode expansion (OME) method \citep{nus95,
dac98}, Wiener filtering method \citep*{zar99}, the unbiased minimal
variance (UMV) estimator \citep{zar02b}, the optimal moment expansion
method \citep{wat02}, and the pairwise velocity method
\citep{fel03}.
Considerable efforts have been made to estimate the $\beta$ parameter
from the comparison of mass density field derived from the observed
radial velocities with the observed galaxy density field from the redshift
survey ($\delta$-$\delta$ comparison; \citealt{sig98}), or from the direct
comparison of the observed peculiar velocity field with that derived
from the galaxy redshift survey
($v$-$v$ comparison; e.g., see \citealt{zar02b}, references therein).

In this paper, we estimate the $\beta$ parameter and the amplitude
of mass fluctuation from the peculiar velocities of the SFI galaxy sample
by measuring the momentum and density power spectra.
The outline of this paper is as follows.
In $\S2$, we briefly review the cosmic momentum field.
Description and data reduction of the SFI galaxy sample are given in $\S3$.
In $\S4$, we measure the density and momentum power spectra from
the SFI data. Measurements of $\beta$ parameter and the amplitude of mass
fluctuation are given in $\S5$. We discuss our results in $\S6$.
Throughout this paper, in calculating the real-space distance from
the redshift or the recession velocity, we assume a flat $\Lambda$CDM
universe with matter and dark energy density parameters
$\Omega_m = 0.27$ and $\Omega_\Lambda = 0.73$, respectively.
For the present value of the Hubble parameter, we use $H_0 = 100 h$ km/s/Mpc.

\section{Cosmic Momentum Field}

This section summarizes the cosmic momentum field.
For complete description we refer to P00.
The cosmic momentum field is defined as the peculiar velocity field $\biv$
weighted by local density $\rho / \bar\rho$:
\begin{equation}
   \bip \equiv {\rho \over \bar\rho} \biv = (1+\delta) \biv.
\end{equation}
Here $\delta = (\rho-\bar\rho)/\bar\rho$ is the dimensionless density contrast.
The cosmic momentum $\bip$ defined here has the dimension of velocity
and is equal to $\biv$ in the linear regime ($|\delta|\ll 1$).

To compare the observed peculiar velocities of galaxies with cosmological
models we measure correlation function (CF) or power spectrum (PS)
of the momentum field.
Suppose the matter over-density $\delta(\bix)$ and the peculiar velocity
$\biv(\bix)$ fields are homogeneous and isotropic random fields.
Given a momentum correlation tensor
$\xi_{ij}^p(r)=\left<p_i(\bix)p_j(\bix+\bir) \right>$,
we define a power spectral tensor $P_{ij}^p(k)$ as
\begin{equation}
   P_{ij}^p(k) \equiv \int d^3 r \xi_{ij}^p (r) e^{i\bik\cdot\bir}.
\end{equation}
With this definition, $P_{ij}^p(k)$ has an unit of
$\textrm{[velocity]}^2 \times \textrm{[volume]}$, conventionally
$\textrm{km}^2\textrm{s}^{-2} (h^{-1}\textrm{Mpc})^3$.
We call its trace
$P_p(k) \equiv P_{ii}^p (k)= \int d^3 x \xi_p (r) e^{i\bik\cdot\bix}$
as the PS of the momentum field, where $\xi_p(r)\equiv\xi_{ii}^p(r)
=\left<\bip(\bix)\cdot\bip(\bix+\bir)\right>$ is the dot-product CF of the
momentum field.

What we directly measure in a galaxy peculiar velocity survey
is the radial component of peculiar velocities at the locations of galaxies.
The radial component of momentum, the physical observable in our analysis,
is the galaxy number-weighted quantity $p_r(\bix) = u(\bix) n(\bix) /\bar{n}$,
where the radial peculiar velocity $u(\bix)$ is caused by the total matter
field and $n(\bix)/\bar{n}$ represents the distribution of galaxies.
With the isotropy of the momentum field and the far-field approximation,
P00 obtained
\begin{equation}
   P_p(k) \approx 3 P_{p_r}(k).
\end{equation}
The three-dimensional momentum PS can be measured from the PS of radial
component of momentum $P_{p_r}(k)$.
This property can be directly applied to the all-sky peculiar velocity
survey data.
By comparing the PS of the total momentum with those of the radial component
of the momentum vector observed at a corner or at the center of the simulation
cube, P00 has demonstrated that equation (3) is actually very accurate
over wide scales when the cosmic variance and observational uncertainties
in the PS are taken into account. However, at scales corresponding
to the fundamental modes within the survey volume, equation (3) holds only
approximately.

If the velocity field is irrotational (curl-free),
Fourier mode of velocity field in the linear regime becomes
\begin{equation}
   \biv(\bik) = -i(DHf) {\bik \over k^2} \tilde\delta(\bik),
\end{equation}
where $D(t)$ is the linear growth factor as in
$\delta(\bik;t) = D(t) \tilde\delta(\bik)$, $H\equiv \dot{a}/a$ is the Hubble
parameter, $a(t)$ is the expansion factor, and
$f(\Omega_m,\Omega_\Lambda) \equiv d\ln D /d\ln a \simeq \Omega_m^{0.6}$
\citep{pee80}. We use the Fourier transform (FT) convention defined as
$\biv(\bik) = (1/V) \int_V d^3 x \biv(\bix) e^{i\bik\cdot\bix}$
on a large volume $V$ over which they are considered to be periodic.
The inverse FT is thus defined as
$\biv(\bix) = V/(2\pi)^3 \int d^3 k \biv(\bik) e^{-i\bik\cdot\bix} 
= \sum_{\bik} \biv(\bik) e^{-i\bik\cdot\bix}$.

From equation (4), the PS of velocity field ($P_v$) in the linear regime
is related with the density PS ($P_\delta$) as
\begin{equation}
    P_v (k) = V \left<|\biv(\bik)|^2\right> = (DHf)^2 P_\delta (k) / k^2.
\end{equation}
From the FT of equation (1),
$\bip(\bik) = \biv(\bik) + \sum_{\bik'}\delta(\bik')\biv(\bik-\bik')$,
the approximate expression for the momentum PS is given by the sum of
PS of $\biv$ and $\delta\biv$ fields (P00):
\begin{equation}
\begin{split}
   P_p&(k) \approx P_v(k) + P_{\delta v}(k)= (DHf)^2 P_\delta (k) /k^2   \\
     &+ {1 \over 2} (D^2Hf)^2 \int {{d^3 k'} \over {(2\pi)^3}} 
     {{k^2} \over {k'^2 |\bik-\bik'|^2}} P_\delta (k') P_\delta (|\bik-\bik'|).
\end{split}
\end{equation}
It should be noted that equation (6) is not a complete expression
for the cosmic momentum PS because only the $\delta\biv$ term is considered
with the other non-linear terms excluded.
Furthermore, equation (6) assumes that $\delta(\bix)$ and
$\biv(\bix)$ are Gaussian with a property that the ensemble of odd-product like
$\left<\delta(\bix)\biv(\bix)\cdot\biv(\bix+\bir)\right>$ is zero.
The correct expression for the momentum PS requires higher order perturbation
theory (e.g., \citealt{ber02}).

\begin{figure}
\mbox{\epsfig{file=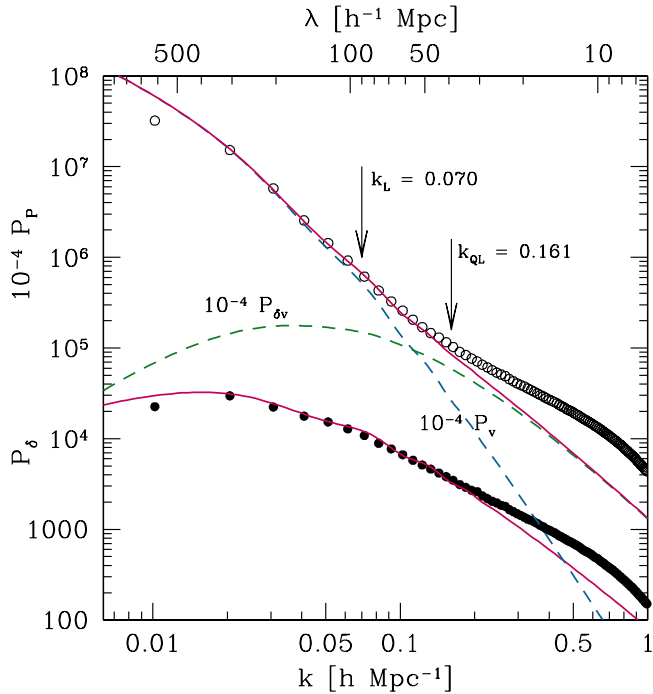,width=88mm,clip=}}
\caption{Momentum and density PS measured from the $N$-body 
simulation of the $\Lambda$CDM universe (open and filled circles), 
in units of $\textrm{km}^2\textrm{s}^{-2} (h^{-1}\textrm{Mpc})^3$ and
$(h^{-1}\textrm{Mpc})^3$, respectively.
The bottom solid curve is the linear matter density PS ($P_\delta$)
for $\Lambda$CDM cosmology as used in the $N$-body simulation.
The upper solid curve denotes the momentum PS ($P_p$) that is the sum of PS
of $\biv$ and $\delta \biv$ fields ($P_v$ and $P_{\delta v}$; dashed curves).
The $P_v$ and $P_{\delta v}$ have been derived from the linear density PS
(eq. [6]) 
The limits of the linear and the quasi-linear regimes, 
denoted as $k_L$ and $k_{QL}$, respectively, are indicated as arrows
(see $\S5$).}
\label{jfig1}
\end{figure}

The accuracy of equation (6) can be tested using the $N$-body simulation data.
Figure 1 shows the momentum and density PS directly measured from a
$N$-body simulation of the $\Lambda$CDM cosmological model (open and filled
circles). A particle-mesh code \citep{par90,par97} is used to gravitationally
evolve $256^3$ CDM particles from $z=23$ to $0$ on a $512^3$ mesh
whose physical size is $614.4$ $h^{-1}$Mpc.
The cosmological parameters used in the simulation are
$\Omega_m = 0.27$, $\Omega_b = 0.0463$, $\Omega_\Lambda = 0.73$, and $h=0.72$,
and the fitting formulae for PS given by \citet{eis99} are used.
The simulation is normalized so that $\sigma_8 = 1/b = 0.9$ at $z=0$, where
$\sigma_8$ is the rms mass fluctuation within 8 $h^{-1}$Mpc sphere.
We use the cloud-in-cell (CIC) scheme (e.g., \citealt{hoc81}) in constructing
density and momentum fields, and the fast Fourier transform (FFT) to estimate
the PS.
The upper solid curve which is the sum of PS of $\biv$ and $\delta\biv$
fields (two dashed curves) is the momentum PS calculated from equation (6)
using the linear matter density PS (bottom solid curve).
Comparison of the momentum PS measured from the $N$-body simulation
(open circles) with the upper solid curve shows that the equation (6)
is very accurate down to scales of about 50 $h^{-1}\textrm{Mpc}$
(at the largest scale corresponding to the simulation
box size, the agreement is not good because of the cosmic variance).
This demonstrates that the momentum PS can be derived from the linear
density PS even in the quasi-linear regime.

\section{The SFI Galaxy Sample}

\subsection{Description of the Sample}

The SFI galaxy catalog is an all-sky sample of peculiar velocity of galaxies.
It contains 1289 Sbc--Sc galaxies with $I$-band TF distances
\citep{gio94,gio97a,gio97b,hay99a,hay99b}.
The galaxies in the catalog have inclination $i\ga 45\degr$ and Galactic
latitudes $|b| > 10\degr$.
Its survey depth is $cz_{\rm LG} = 7500$ km/s, where $cz_{\rm LG}$ is the
galaxy recession velocity with respect to the Local Group (LG).
This sample is a combination of the SFI main galaxy sample
($\delta > -45\degr$) and \citet[hereafter MAT]{mat92} sample of which
the magnitudes and rotational velocities were converted to the SFI system.

\begin{figure}
\mbox{\epsfig{file=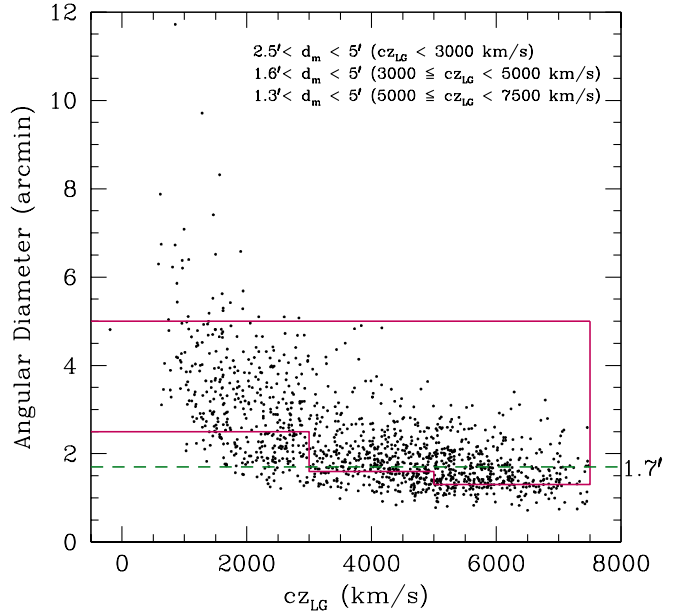,width=88mm,clip=}}
\caption{Redshift ($cz_{\rm LG}$) versus angular diameter ($d_m$) of
1289 galaxies in the SFI catalog. The $d_m$ is obtained from a relation
$d_m/a_{23.5} = 2.78-1.03 \log(a/b)$, where $a_{23.5}$ is the isophotal
angular radius measured at $\mu_I=23.5$ mag/arcsec$^2$ level
and $a/b$ is the axial ratio.
The region enclosed by solid lines represents the SFI main galaxy
selection criteria, while the dashed line denotes our angular diameter limit
applied to the second and the third redshift shells.}
\label{jfig2}
\end{figure}

The peculiar velocities of spiral galaxies in the SFI sample have been derived
from the $I$-band TF relation.
We consider the TF relation between the absolute magnitude $M_I$ and
the velocity width parameter $\eta$,
\begin{equation}
   M_I = m_I - 5\log r = a_{\rm TF} + b_{\rm TF} \eta,
\end{equation}
where $\eta \equiv \log w - 2.5$ and $w$ is the circular velocity line-width
of a galaxy in unit of km/s. The absolute magnitude $M_I$ is defined as
the apparent magnitude when a galaxy is located at $r=1$ km/s.
For the slope and the zero-point of the
TF relation, we adopt $a_{\rm TF} = -21.10$ and $b_{\rm TF}=-7.94$
of the inverse Tully-Fisher (ITF) relation that is determined by \citet{gio97a}
for 24 clusters in the SCI sample.
The reason for using the ITF relation is worth mentioning.
The forward TF relation is obtained by regressing the apparent
magnitudes over the line-width. It can be biased due to the imposed selection
limits on magnitude, angular diameter, and circular velocity line-width.
On the other hand, the ITF relation is obtained by fitting the line-width
as a function of the apparent magnitude. It avoids the selection bias
if the sample selection is independent of the line-width
\citep{str95,fre95,bor00a}.
We perform our analysis using the peculiar velocities inferred from the ITF
relation. \citet{bor00a} analyzes the SFI peculiar velocity data
in redshift space to avoid the possible bias that can enter
into the measurement of distances or peculiar velocities of galaxies.
However, we make our analysis in {\it real} space to avoid the effects
from the redshift space distortion.

\begin{figure}
\mbox{\epsfig{file=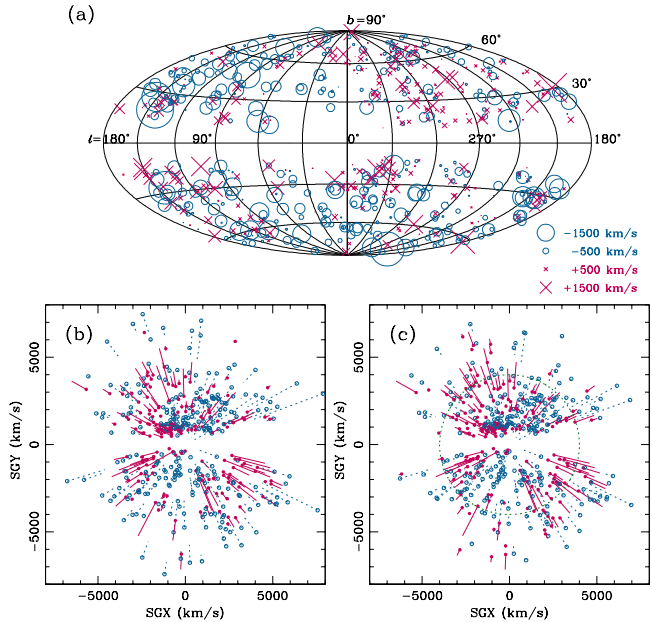,width=88mm,clip=}}
\caption{($a$) Hammer-Aitoff sky projection in the Galactic coordinate as seen
in the LG-frame, of the 661 SFI galaxies. Open circles indicate
the infall while crosses outflow. The size of symbols scales with the
velocity amplitude as shown in the bottom-right of the figure.
($b$) \& ($c$) Distribution of the SFI galaxies in the supergalactic
(SG) coordinates before and after MB correction, respectively. 
The positions and peculiar velocities are shown in the  
$\textrm{SGX}$--$\textrm{SGY}$ plane with thickness of $\Delta{\rm Z} < 2000$
km/s. 
Outflowing galaxies are denoted as dots with solid segments that scale with
the velocity amplitude, while infalling galaxies as open circles with dotted
segments. The large circle in panel ($c$) represents a region enclosed by
a $40~h^{-1}\textrm{Mpc}$ sphere.}
\label{jfig3}
\end{figure}

The SFI galaxy sample, by desire, is angular-diameter
limited. The sample was made to obtain roughly the same number of galaxies
in three different redshift shells. Diameter limits imposed on the SFI
main galaxy sample on each redshift shell are
$2\farcm5 < d_m < 5\arcmin$ at $cz_{\rm LG} < 3000$ km/s,
$1\farcm6 < d_m < 5\arcmin$ at $3000 \leq cz_{\rm LG} < 5000$ km/s, and
$1\farcm3 < d_m < 5\arcmin$ at $5000 \leq cz_{\rm LG} < 7500$ km/s.
Here $d_m$ is the angular diameter defined as the UGC blue major axis
or the analogous parameter in the ESO-Uppsala Survey \citep{gio94}.
The distribution of SFI galaxies on a $cz_{\rm LG}$--$d_m$ plane is shown
in Figure 2. The galaxies actually do not strictly follow the selection
criteria.
The angular diameter of each galaxy is estimated from a relation
$d_m/a_{23.5}=2.78 - 1.03 \log(a/b)$, where $a_{23.5}$ is
the $I$-band isophotal radius measured at $\mu_I=23.5$ mag/arcsec$^2$ level
and $a/b$ is the axial ratio of a galaxy \citep{gio94}.
The angular diameter limit of the MAT galaxy sample is $1\farcm7$.
For completeness of the SFI sample, we exclude galaxies with diameter
smaller than $1\farcm7$ on the second and the third redshift shells.
A small fraction of galaxies with small line-widths ($\log w < 2.25$)
have also been discarded because of the large fractional errors
in the measurement of their widths and thus the large uncertainties
in the peculiar velocity measurement \citep{dac96}.
The total number of galaxies we use in our analysis
is 661. Figure 3 shows the positions and peculiar velocities of the 661 SFI
galaxies on the sky and the supergalactic plane.
The supergalactic coordinate system is a Cartesian coordinate system
with the $\textrm{SGZ}$-axis pointing to the Galactic coordinate $l=47\fdg37$,
$b=6\fdg32$, and the $\textrm{SGX}$-axis pointing to $l=137\fdg29$, $b=0\degr$
(e.g., \citealt{pee93}).

\subsection{Bias Correction}
Real data are often selected by magnitude or diameter limits, and/or by
redshift limit, which cause a bias in the calibration of the TF relation.
This is particularly important for the SFI sample because
the adopted redshift-dependent selection criteria imply that near each
redshift limit outflowing galaxies are preferentially excluded from the sample.
As shown in Figure 3$b$ this effect becomes dominant at large distances,
leading to a spurious systematic infall of galaxies at the outer edge of
the survey volume \citep{dac96}.
In earlier studies with the SFI data, the biases were estimated
using the numerical Monte-Carlo technique \citep{fre95,dac96,bra01}
or using the semi-analytic approach \citep{fre99}.
In this paper, we incorporate all the selection effects into two selection
functions:
a radial function $\phi(r)$ for the redshift-dependent selection,
and an angular function $\psi(b)$ for the Galactic extinction
that may reduce the galaxy number density at the low Galactic latitude.
For the radial selection function we use a method that is equivalent
to the $V/V_{\rm max}$ method for computing the luminosity function
\citep{sch68}. The radial selection function $\phi(r)$ is given by
\citep{par94,str95}
\begin{equation}
   \phi(r) = {3 \over \Omega_s} \sum_{d_{{\rm max},i}>r} 
             {{1}\over{d_{{\rm max},i}^3}},
\end{equation}
where $\Omega_s$ is the solid angle covered by the survey and
$d_{{\rm max},i}$ is the maximum distance the $i$-th galaxy can have
while satisfying the selection criteria.

Malmquist bias is caused by the random error in the galaxy distances
estimated by the distance estimators like TF or $D_n$--$\sigma$ relations
\citep{lyn88}.
At a given distance there are more galaxies that are perturbed from
the far side than from the near side.
The number of galaxies that are randomly moved to that distance
also depends on the actual distribution of galaxies.
Biases in the inferred distance or the peculiar velocity occur
because of the volume effect and galaxy number density variation
along the line of sight.
We call the combination of the two effects the Malmquist bias (hereafter MB).
To correct for MB, we have derived a density distribution of
the local universe using the IRAS PSC$z$ 0.6 Jy flux-limited catalog
\citep{sau00}.
We assume the linear theory of gravitational instability to estimate
the peculiar velocity of each PSC$z$ galaxy, and obtain the real-space
density field using the method described in \citet{bra99}.
In deriving the density field we assume $\beta = 0.5$.
Given a TF distance ($d$) of each galaxy, we obtain MB-corrected
distance using a relation \citep{str95}
\begin{equation}
   E(r|d) = {{\int_0^\infty r^3 n(r) \phi(r) 
            \exp \left( - [\ln(r/d)]^2 / 2\Delta^2 \right) dr}
            \over {\int_0^\infty r^2 n(r) \phi(r)
            \exp \left( - [\ln(r/d)]^2 / 2\Delta^2 \right) dr}}.
\end{equation}
Here $n(r)$ is the number density of galaxies along the line-of-sight,
$r$ denotes unbiased true distance, and
$\Delta=\sigma_{\rm TF}\ln10/5$ is a measure of the fractional distance
uncertainty, where $\sigma_{\rm TF}$ is the error of the TF relation
in unit of magnitude.
In this paper, we assume $\sigma_{\rm TF} = 0.36$ \citep{gio97a}.
The $\phi(r)$ in equation (9) is the real-space radial selection function
derived from the raw SFI sample ({\it not} corrected for MB;
dotted curve in Fig. 4).
Given the fully corrected distance estimate, $d_c = E(r|d)$,
the radial component of the peculiar velocity is obtained as
$u = cz_{\rm LG} - d_c$. The computed distances and peculiar velocities
are used to measure the cosmic density and momentum PS ($\S4.3$).

\begin{figure}
\mbox{\epsfig{file=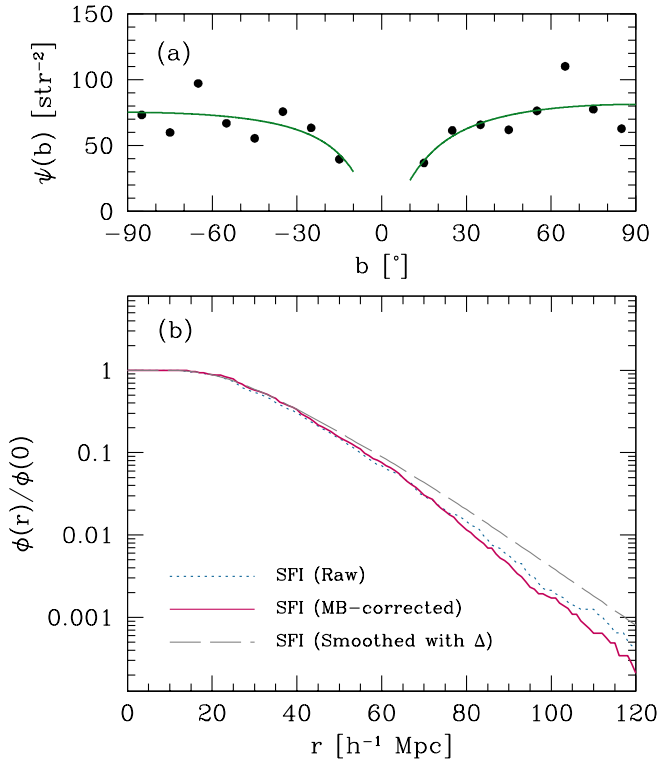,width=88mm,clip=}}
\caption{Angular (upper) and radial (bottom) selection functions
of the 661 galaxies in the SFI sample. For the angular selection function
we estimate the number density of galaxies at Galactic latitude bins
with $\Delta b = 10\degr$.
The estimated number densities have been fitted with a function
$\psi(b) = \psi_0 10^{\alpha (1-\csc|b|)}$, where $\psi_0 = 81.3$ ($75.2$)
and $\alpha = 0.113$ ($0.084$) for the northern (southern) hemisphere.
Radial selection functions before and after MB correction, and that
further smoothed with $\Delta = \sigma_{\rm TF}\ln10 /5$ are shown
as dotted, solid, and dashed curves, respectively.}
\label{jfig4}
\end{figure}

Figure 4 shows the angular and radial selection functions of the 661 galaxies
in the SFI catalog.
The angular selection function ($\psi$) is obtained by fitting the galaxy
number density estimates at Galactic latitude bins to a function
$\psi(b)=\psi_0 e^{\alpha (1-\csc|b|)}$.
We calculate radial selection functions ($\phi$) from the 661 galaxies
before and after MB correction (dotted and solid
curves, respectively), and MB-corrected selection function smoothed
with a Gaussian filter of width $\Delta$ (dashed curve).
The final selection function is given by $s(r,b) = \phi(r) \psi(b)$.

\subsection{Survey Window Function}

\begin{figure}
\mbox{\epsfig{file=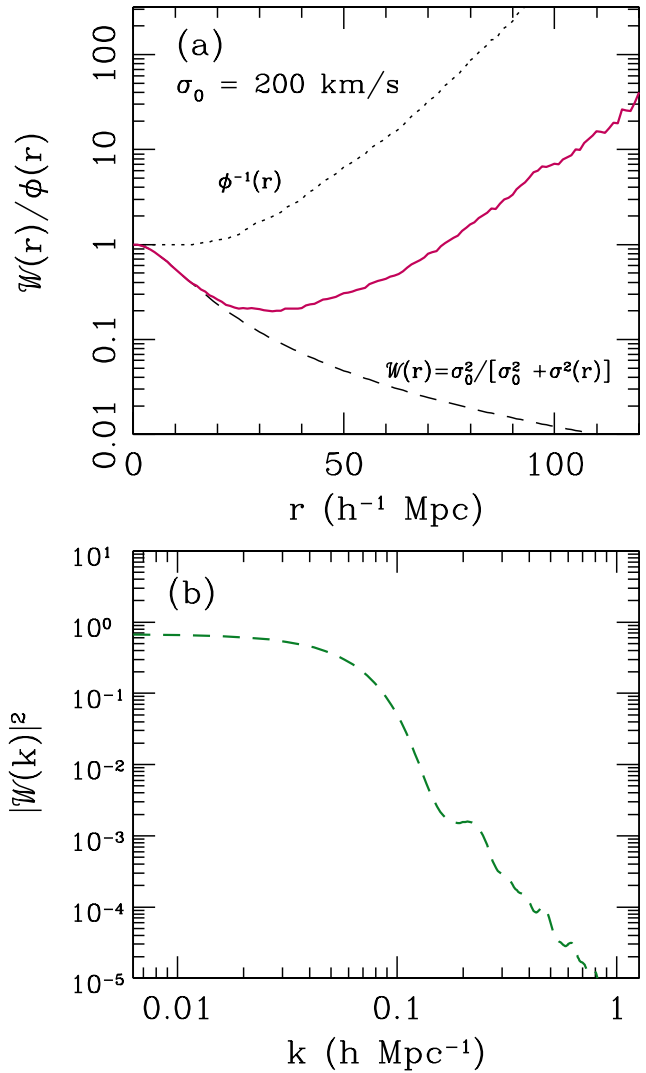,width=88mm,clip=}}
\caption{(Upper) Effective weight of a galaxy at distance $r$ (solid curve),
shown as the product of the inverse of the radial selection function
($\phi^{-1}(r)$; dotted) and the survey window function ($\MW(r)$; 
dashed curve) for $\sigma_0 = 200$ km/s and $\sigma_{\rm TF}=0.36$.
Here $\phi(r)$ is normalized so that $\phi(0)=1$. 
(Bottom) PS of the survey window function $|\MW(\bik)|^2$. The survey region
is limited to a volume within a sphere of $r_{\rm max}=40$ $h^{-1}$Mpc
and $|b| > 10\degr$.}
\label{jfig5}
\end{figure}

Although galaxy positions in redshift space are measured very
accurately, the error in the absolute distances and the peculiar velocities
of galaxies monotonically increases radially outward, and dominates the
signal beyond a certain distance.
The expected distance error of a galaxy at distance $r$ (in $h^{-1}$Mpc)
is approximately $\sigma(r) = 100 r (e^\Delta - 1)$ km/s.
To reduce the effects of the large distance error on the measured PS,
we use a survey window function defined as
\begin{equation}
   \MW(r) = {{\sigma_0^2} \over {\sigma_0^2 + \sigma^2 (r)}}. 
\end{equation}
With the smaller $\sigma_0$, the effective survey depth in our analysis
becomes shallower.
Figure 5 shows the inverse of radial selection function $\phi^{-1}(r)$
and the window function $\MW(r)$ (dotted and dashed curves).
The product $\phi^{-1}(r)\MW(r)$, the solid curve in Figure 5$a$,
is the effective weight of a galaxy at distance $r$ (see eq. [12] in $\S4.1$).
We adopt $\sigma_0 = 200$ km/s and limit the sample depth to
$r_{\rm max}=40~h^{-1}\textrm{Mpc}$ to exclude distant galaxies with
large effective weights.
The PS of the survey window function shown in Figure 5$b$ is a monotonically
and rapidly decreasing function, and is negligible at
$k > 0.1~h\textrm{Mpc}^{-1}$. That is, the measured PS has correlations
only within about $\Delta k = 0.1~h\textrm{Mpc}^{-1}$.

\section{Power Spectrum Estimation}

\subsection{Method}

In this section we describe the method of estimating the PS of 
the momentum field $P_p(k)$ from the observed radial peculiar velocity data.
We use the PS estimation method developed by P00 (see also \citealt{par94}).
The method uses a direct FT to calculate the Fourier modes of density 
and momentum fields. In this paper we slightly modify the direct FT method
so that it can be applied to a case for an arbitrary window function.

Suppose we have $N$ galaxies with positions $\bix_j$ and radial peculiar
velocities $u(\bix_j)$, with $j=1,\ldots,N$. Owing to the variation of
selection, each galaxy has a different statistical weight
that is given by the inverse of the selection function, 
$w_j=s^{-1}(\bix_j)$, where the sample selection function $s(\bix_j)$
is defined to be the fraction of galaxies at position $\bix_j$ expected
to be observable by the survey. 
If the galaxies represent mass, the density contrast is related to
the galaxy number density $n(\bix)$ as in
\begin{equation}
   1+\delta(\bix) = {n(\bix) \over {\bar n}} = {1 \over {\sum_j w_j / V}} 
                      \sum_j w_j \delta^{\rm (3)} (\bix - \bix_j),
\end{equation}
where $\delta^{(3)} (\bix-\bix_j)$ is the Dirac delta function,
and $V$ is the survey volume.

It is quite reasonable to assume that $\left< \biv \right> = 0$
for a sufficiently large volume. 
We estimate the mean peculiar velocity $\biv_0$ using a bulk flow model
for the radial peculiar velocities within $r_{\rm max}$.
The mean velocity $\biv_0$ is obtained by minimizing 
$\chi^2 = \sum_j \MW_j^2 (u_j - \biv_0\cdot\hat{\bix}_j)^2$,
where $\MW_j=\MW(\bix_j)$ and $u_j = u(\bix_j)$, and 
$\hat{\bix}_j$ is an unit vector pointing to the $j$-th galaxy \citep{dac00a}.
Then, we obtain a new radial peculiar velocity 
of each galaxy by subtracting a line-of-sight component of the mean peculiar
velocity from the galaxy's observed radial peculiar velocity:
$v_r(\bix_j)=u(\bix_j)-\biv_0\cdot\hat{\bix}_j$.

The FT of the observed radial momentum field is
\begin{equation}
\begin{split}
   \hat{p}_r(\bik) &= {1 \over V}\int d^3 x \MW(\bix) 
      \left[\left(1+\delta(\bix)\right) \biv(\bix)\cdot \hat{\bix} \right] 
      e^{i\bik\cdot\bix}  \\
      &= {1 \over {\sum_j w_j}} \sum_{j=1}^N \MW(\bix_j) w_j v_r(\bix_j) 
        e^{i\bik\cdot\bix_j}.
\end{split}
\end{equation}
Likewise, the FT of $\delta(\bix)$ is given by 
$\hat{\delta}(\bik)=\sum_j \MW(\bix_j)w_je^{i\bik\cdot\bix_j}
/\sum_j w_j-\MW(\bik)$, where $\MW(\bik)$ is the FT of the survey window
function.
The quantities with a caret, $\hat{p}_r(\bik)$ and $\hat{\delta}(\bik)$,
are the momentum and density Fourier modes that are convolved
with the survey window function.
The ensemble average of $|\hat{p}_r(\bik)|^2$ is given by
\begin{equation}
\begin{split}
    &\left(\Sigma_j w_j \right)^2 \left < |\hat{p}_r(\bik)|^2 \right>  \\
     &= \left< \sum_{j\in V} \MW_j w_j v_{rj} e^{i\bik\cdot\bix_j} 
          \sum_{m\in V}\MW_m w_m v_{rm} e^{-i\bik\cdot\bix_m} \right> \\
     &= \sum_{j {\rm (cells)}} \sum_{m {\rm (cells)}} \MW_j \MW_m w_j w_m
     \left< n_j n_m v_{rj} v_{rm} \right> e^{i\bik\cdot (\bix_j-\bix_m)},
\end{split}
\end{equation}
where $v_{rj} = v_r(\bix_j)$.
In the second equality, we divide the survey volume $V$ into infinitesimal
cells with occupation number $n_j =0$ or $1$.
The ensemble averaged quantity
$\left<n_j n_m v_{rj} v_{rm}\right>$ is related with the radial momentum
CF ($\xi_{p_r}$) as
\begin{equation}
\begin{split}
   &\left< n_j n_m v_{rj} v_{rm} \right> \\
   &= \left\{ \begin{array}{ll} 
          \bar{n} s_j d^3x_j \xi_{p_r}(0)  
          & \textrm{if $j=m$} \\
          \bar{n}^2 s_j s_m d^3x_j d^3x_m \xi_{p_r}(|\bix_j-\bix_m|)
          & \textrm{if $j \ne m$} 
            \end{array} \right.,
\end{split}
\end{equation}
where $s_j = s(\bix_j)$, and we use a property
$\left< n_j \right> = \bar{n} s_j d^3 x_j$.
Similarly, for density PS we need to consider $\left<n_j n_m\right>$.
The right-hand side of equation (14) becomes
$\sum_{j}w_j^2 \MW_j^2 \xi_{p_r}(0)$ for $j=m$ case, and
$(\sum_j w_j)^2 \sum_{\bik'} |\MW(\bik')|^2 \left<|p_r(\bik-\bik')|^2\right>$
for $j \ne m$ case.
The final formulae for the density and momentum field PS are
\begin{equation}
   P_{\delta} (k)\approx V \left [ \left\langle |\hat{\delta}(\bik)|^2 
      \right\rangle - {{\sum_j w_j^2 \MW_j^2} \over {(\sum_j w_j})^2} \right ] 
         \bigg/ \sum_{\bik} |\MW(\bik)|^2, 
\end{equation}
and
\begin{equation}
   P_p (k)\approx 3 V \left [ \left\langle |\hat{p}_r(\bik)|^2 \right\rangle  
     - {{\sum_j w_j^2 \MW_j^2} \over {(\sum_j w_j})^2} 
       \xi_{p_r}(0) \right ] \bigg/ \sum_{\bik} |\MW(\bik)|^2, 
\end{equation}
where the radial momentum CF at the zero lag can be estimated as
\begin{equation}
   \xi_{p_r}(0) = {{\sum_j w_j^2 \MW_j^2 v_{rj}^2}\Big/{\sum_j w_j^2 \MW_j^2}}.
\end{equation}
In equations (15) and (16), we assume that PS we are measuring are slowly
varying functions, and approximate that the window-convolved PS ($\hat{P}$)
can be separated into the true PS part ($P$) and the window-related part as
$\hat{P}(k)\approx P(k) \sum_{\bik}|\MW(\bik)|^2$.
The $\sum_{\bik}|\MW(\bik)|^2$ in the denominator is the overall power
correction that compensates for power loss due to the finite window function
that decreases radially outward.

\subsection{Momentum Power Spectrum from Mock Surveys}

We use the $N$-body simulation data in $\S2$ to make mock SFI observations.
To mimic the LG environment, we select an observer as a particle with total
velocity 600--700 km/s on the over-dense region with $0 < \delta < 1$.
From the chosen observer we randomly select the particles
with probability given by $s(r,b)$.
The radial selection is made at the true distance using the radial selection
function derived from the MB-corrected SFI galaxy distribution
(solid curve in Fig. 4).
After the particle selection, we perturb the particle distances with Gaussian
errors of $\Delta$.
By applying redshift cut at $cz_{\rm LG}=7500$ km/s we finally obtain 661
angular positions and recession and peculiar velocities of `galaxy' particles
for each mock observation.
For PS measurement, we use equation (9) to obtain the MB-corrected galaxy
distances and peculiar velocities. Here we have smoothed the $N$-body
simulation data to obtain $n(r)$ by applying a Gaussian filter
with $\sigma=5$ $h^{-1}$Mpc, and used the radial selection function
smoothed with $\Delta$ (dashed curve in Fig. 4).

\begin{figure}
\mbox{\epsfig{file=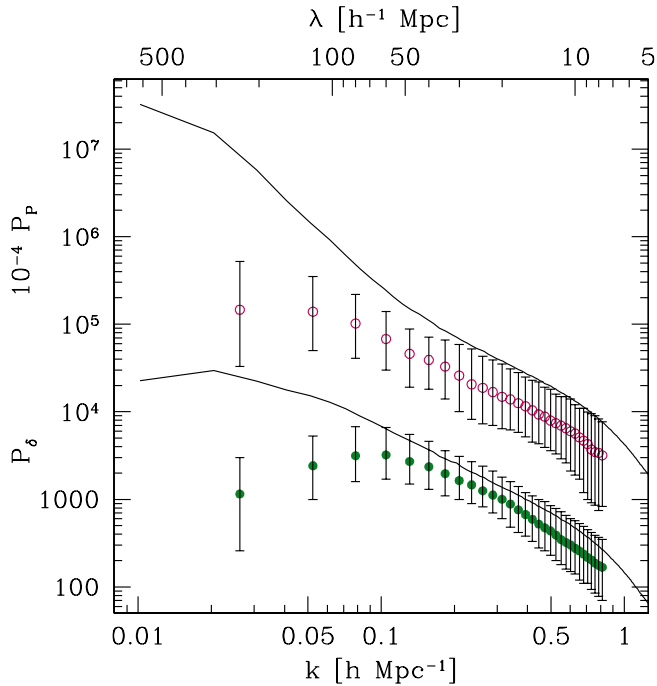,width=88mm,clip=}}
\caption{Momentum and density PS measured from 500 SFI mock 
observations. Shown are the median values (open and filled circles)
and 68\% confidence limits of the PS in the $\Lambda$CDM model.
The solid curves are the true PS obtained from the $N$-body simulation data.}
\label{jfig6}
\end{figure}

The PS of momentum and density fields measured from the five hundred SFI
mock observations are shown in Figure 6. For each mock data
the number of galaxies within $r_{\rm max}=40$ $h^{-1}$Mpc
varies from 300 to 400.
Shown are the median values (open and filled circles) and 68\% confidence
limits of $P_p(k)$ (i.e., $3P_{p_r}(k)$) and $P_\delta(k)$ of the 500 SFI mock
PS. Note that data points in the PS are correlated with each
other over approximately four neighboring points ($\Delta k \simeq 0.1$
$h\textrm{Mpc}^{-1}$).
The solid curves are the true PS of momentum (upper) and density (bottom)
fields of the $\Lambda$CDM universe measured from the $N$-body simulation
(same as open and filled circles in Fig. 1).
The damping of powers at large scales (or small $k$) occurs due to
finiteness of the survey volume. Similar damping is also seen at high $k$,
which is due to the smoothing effect of the peculiar velocity and distance
errors.
We use the ratio of the true PS to the PS of mock surveys as the correction
factor when we measure the PS for the observed SFI sample.
At each $k$, the raw power measured from the observed data is multiplied
with the corresponding the correction factor. Then, the systematic effects
in the raw powers are removed, and the PS with correct amplitudes
can be restored. Those factors also reduce any residual biases.
This method of correction for the systematic effects in the observed galaxy
PS by using mock surveys has been developed by \citet{par92,par94} and
\citet{vog92}.

\subsection{Application to the SFI Data}

\begin{figure}
\mbox{\epsfig{file=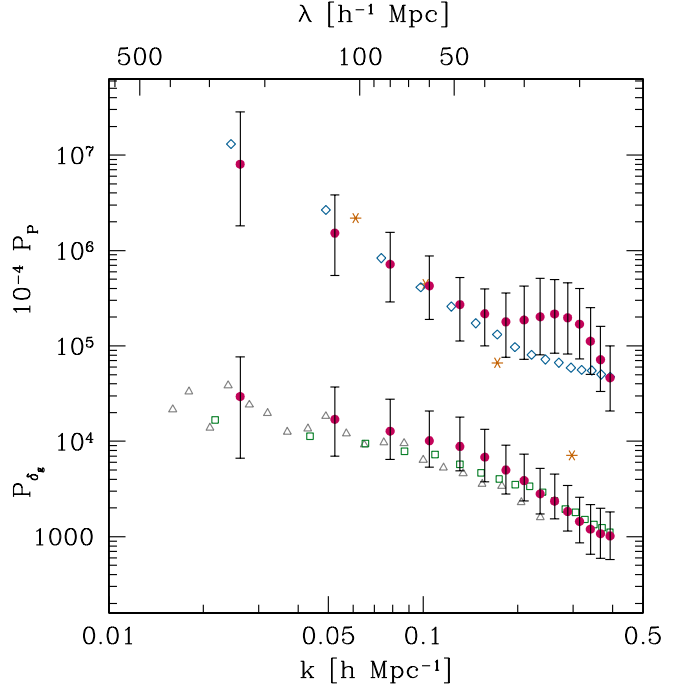,width=88mm,clip=}}
\caption{Momentum and density PS measured from the SFI galaxy sample
(filled circles). 
The velocity PS measured from the Mark III catalog with POTENT method
is shown as stars \citep{kol97}, and the momentum PS measured by P00 for the
MAT data as diamonds.
Open triangles denote the decorrelated real-space galaxy density PS
measured from the SDSS galaxy redshift survey \citep{teg04}.
Open squares represent the galaxy density PS in the real space calculated
by \citet{par94} for the 101 $h^{-1}$Mpc deep subsample of the CfA survey.}
\label{jfig7}
\end{figure}

We analyze the SFI catalog using the same method as applied to the mock
SFI catalogs. The number of galaxies used in the analysis is 370
within $r_{\rm max}=40~h^{-1}\textrm{Mpc}$.
The final PS of the observed SFI galaxies obtained by multiplying
the correction factor to the measured power spectrum at each $k$
are shown in Figure 7 (filled circles).
The PS are measured in the true distance space rather than in the redshift
space. The 68\% uncertainty limit at each wavenumber is estimated from the five
hundred mock surveys in the $\Lambda$CDM model.
The Poisson noise powers for the density and momentum PS are
$523~(h^{-1}\textrm{Mpc})^3$ and
$7.5\times 10^8~\textrm{km}^2\textrm{s}^{-2}(h^{-1}\textrm{Mpc})^3$,
respectively.
For comparisons, we plot the velocity PS measured by \citet{kol97}
who applied the POTENT method to the Mark III catalog (stars), and
the momentum PS measured by P00 for the MAT data (diamonds).
The velocity field is close to linear at scales
$k \lesssim 0.07$ $h$Mpc$^{-1}$, where the momentum PS should be
equal to the velocity PS (P00).
We find that the velocity PS of \citeauthor{kol97}
is in good agreement with the SFI momentum PS only at the first two wavenumbers,
while the MAT momentum PS is consistent with the SFI PS at all wavenumbers.

We also plot the real-space galaxy density PS for the Sloan Digital Sky Survey
(SDSS; open triangles; \citealt{teg04})
and the Center for Astrophysics (CfA; open squares; \citealt{par94})
galaxy samples.
The measured SFI density PS is in good agreement with the SDSS and
the CfA density PS even though the size of the SFI survey volume and the number
of galaxies analyzed are much smaller.
Compared to the SDSS and the CfA PS, the SFI density PS
has relatively higher power at scales around 40 $h^{-1}$Mpc.
The high power at those scales, considered to be a sampling
effect due to the small survey volume, induces somewhat large value of
$\sigma_{8,\rm S}=1.14_{-0.34}^{+0.33}$,
the rms mass fluctuation of the spiral galaxies within 8 $h^{-1}$Mpc sphere.
The subscript ${\rm S}$ denotes the optical spiral galaxies.
The 68\% uncertainty limits are from the 500 mock surveys.

\section{Estimating the $\beta$ Parameters}

Until now, we have assumed that the galaxy distribution represents the mass
field. If this is not true, the formulae containing the $\delta$ term
should be modified because the observed momentum is now $(1+\delta_g)\biv$.
Suppose galaxies are linearly biased with respect to mass by a constant factor,
i.e., $\delta_g=b\delta$, where $b$ is the linear bias factor of galaxy
distribution.
In the linear regime the PS of velocity field in equation (5) becomes
\begin{equation}
   P_v(k) = (H_0\beta)^2 P_{\delta_g}(k)/k^2,
\end{equation}
where $\beta = f(\Omega_m,\Omega_\Lambda)/b \simeq \Omega_m^{0.6}/b$
and $P_{\delta_g}$ is the galaxy density PS.
P00 has proved that only the overall amplitude of the momentum PS given
by equation (6) is scaled by a factor $\beta^2=(f/b)^2$ due to biasing
if $\beta$ is independent of $k$.

Since we calculate both density and momentum PS from the same sample,
it is possible to measure the $\beta$ parameter more accurately.
We estimate $\beta_{\rm S}=\Omega_m^{0.6}/b_{\rm S}$ parameter
from the measured SFI PS.
As observed by P00, there is a fair amount of correlation between
the estimated $P_\delta$ and $P_p$.
This is because density and momentum PS measured
from the same sample tend to fluctuate statistically in a similar way.
The correlation makes their ratios less uncertain, making the estimated
$\beta$ parameter more accurate.
Furthermore, we note that the momentum PS can be derived from the observed
density PS even in the quasi-linear regime ($\S1$, Fig. 1).
We estimate $\beta_{\rm S}$ from the following formula
\begin{equation}
   \beta_{\rm S}(k) = {{P_p^{\rm obs}(k)} \over {P_p^{\rm der}(k)}}.
\end{equation}
Here the derived momentum PS, $P_p^{\rm der}$, is calculated by equation (6)
with the density PS replaced by $P_{\delta_g}^{\rm obs}$ and $D=f=1$.
During the numerical integration, we interpolate the density PS between
the measured data points, and for small $k \lesssim 0.02~h\textrm{Mpc}^{-1}$
we extrapolate the PS using the $\Lambda$CDM linear density PS
that has been scaled to match the observed PS.
For large $k \gtrsim 1~h\textrm{Mpc}^{-1}$ we also extrapolate
the PS by a power-law function that fits the observed PS at high $k$.
Figure 8 shows the estimated $\beta_{\rm S}(k)$ at each wavenumber $k$ with
68\% limits including the cosmic variance (filled and open circles
with error bars).
The uncertainties have been determined from the distribution of
$\beta_{\rm S}$ parameters of the 500 mock observations, where
the PS of each mock observation are scaled to give the observed SFI PS
on average.

\begin{figure}
\mbox{\epsfig{file=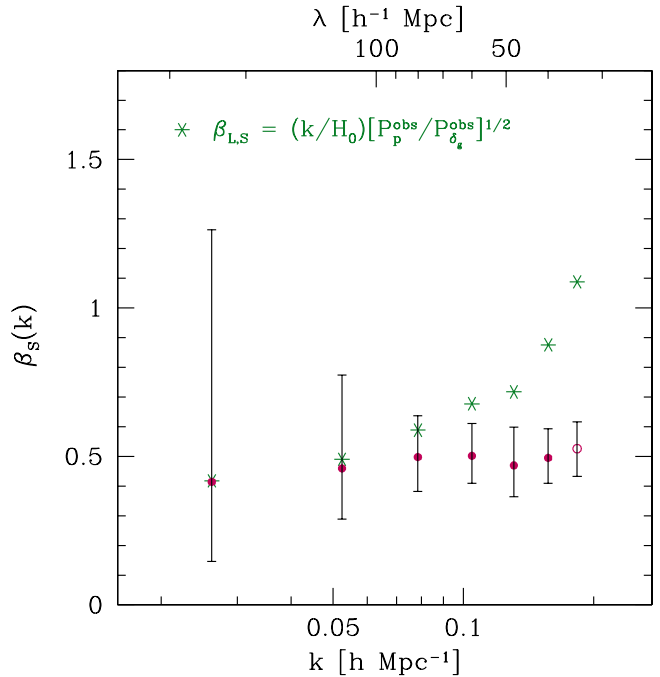,width=88mm,clip=}}
\caption{$\beta_{\rm S}=\Omega_m^{0.6}/b_{\rm S}$ parameters estimated from equation (19)
(filled and open circles), together with $\beta_{L,\rm S}(k)
=(k/H_0)[P_p^{\rm obs}(k)/P_{\delta_g}^{\rm obs}(k)]^{1/2}$ (stars).
The uncertainties represents the 68\% confidence limits which include the
cosmic variance. The first six data points (filled circles) have been used
to obtain the weighted average of $\beta_{\rm S}$ parameters (see Table 1).}
\label{jfig8}
\end{figure}

\begin{deluxetable*}{ccccc}
\tablewidth{350pt}
\tabletypesize{\small}
\tablecaption{$\beta_{\rm S}=\Omega_m^{0.6}/b_{\rm S}$ and
  $\Omega_m^{1.2}P_{\delta}$ Measured from the SFI Peculiar Velocity Sample}
\tablecomments{$^\dagger$$\beta_{L,\rm S}(k)
   =(k/H_0)[P_p^{\rm obs}(k)/P_{\delta_g}^{\rm obs}(k)]^{1/2}$}
\tablecolumns{5}
\tablehead{ $k$ ($h$Mpc$^{-1}$)  & $\lambda$ ($h^{-1}$Mpc) &
            $\beta_{L,\rm S}(k)^\dagger$ & $\beta_{\rm S}(k)$  &
            $\Omega_m^{1.2}P_{\delta}(k)$ $(h^{-1}\textrm{Mpc})^3$ }
\startdata
 $0.0262$ & $240$ & $0.42_{-0.20}^{+0.66}$ & $0.414_{-0.198}^{+0.629}$ 
    & $4.81_{-4.00}^{+9.95}\times 10^3$\\[1mm]
 $0.0524$ & $120$ & $0.49_{-0.16}^{+0.30}$ & $0.459_{-0.147}^{+0.272}$
    & $3.50_{-1.71}^{+3.68}\times 10^3$\\[1mm]
 $0.0785$ &  $80$ & $0.59_{-0.13}^{+0.19}$ & $0.498_{-0.106}^{+0.128}$
    & $3.11_{-1.06}^{+1.58}\times 10^3$\\[1mm]
 $0.1047$ &  $60$ & \nodata& $0.502_{-0.087}^{+0.103}$ 
    & $2.51_{-0.94}^{+0.91}\times 10^3$\\[1mm]
 $0.1309$ &  $48$ & \nodata& $0.469_{-0.099}^{+0.121}$ 
    & $1.91_{-0.78}^{+1.02}\times 10^3$\\[1mm]
 $0.1517$ &  $40$ & \nodata& $0.495_{-0.083}^{+0.096}$
    & $1.65_{-0.56}^{+0.75}\times 10^3$ \\[1mm]
\cline{1-5}
\\[-1mm] \multicolumn{2}{c}{Average} &
  $0.55_{-0.14}^{+0.29}$ &
  $0.489_{-0.050}^{+0.080}$  & \nodata \\[-1mm] 
\enddata
\end{deluxetable*}

We determine limits of the linear and quasi-linear regimes
($k_L$ and $k_{QL}$) by comparing the true momentum PS
with the velocity and momentum PS derived from the linear density PS
(see Fig. 1). The $k_L$ ($k_{QL}$) is defined as the scale at which the derived
velocity (momentum) PS deviates from the true momentum PS (open circles
in Fig. 1) by 20\% in power.
The two limits are $k_L = 0.070$ $h$Mpc$^{-1}$ and $k_{QL} = 0.161$
$h$Mpc$^{-1}$, as denoted by arrows in Figure 1, and
correspond to wavelengths of $90$ and $39$ $h^{-1}$Mpc, respectively.
Table 1 lists the $\beta$ parameters measured from the SFI galaxy
sample in the range $0.0262$ $h$Mpc$^{-1}$ $\le k \le 0.1571$ $h$Mpc$^{-1}$
(six filled circles in Fig. 8).
The weighted average is $\beta_{\rm S}= 0.489_{-0.050}^{+0.080}$.
We call $\beta_{L,\rm S}$ as $\beta$ parameter calculated by using
equation (18) in the linear scales $k \lesssim k_L$.
They are also listed in Table 1, with a weighted average
$\beta_{L,\rm S}=0.55_{-0.14}^{+0.29}$.
In estimating $\beta_{L,\rm S}$, we include $\beta$ at 80 $h^{-1}$Mpc scale
($k=0.0785$ $h$Mpc$^{-1}$) where its uncertainty dominates over
the 20\% difference in power between the velocity and momentum PS.
The $\beta_{L,\rm S}$ is very uncertain because the information contained
in the survey volume is not enough to constrain the PS to an accurate value
at large scales. It starts to deviate from $\beta_{\rm S}(k)$ significantly
at $k > k_L$ where the linear gravitational instability theory no longer holds
(see the stars in Fig. 8).
In Figure 9 the observed SFI momentum PS (upper open circles)
is compared with that derived from the observed density PS (diamonds)
when $\beta=0.489$. The momentum PS prediction is accurate
on scales down to 40 $h^{-1}$Mpc.

\begin{figure}
\mbox{\epsfig{file=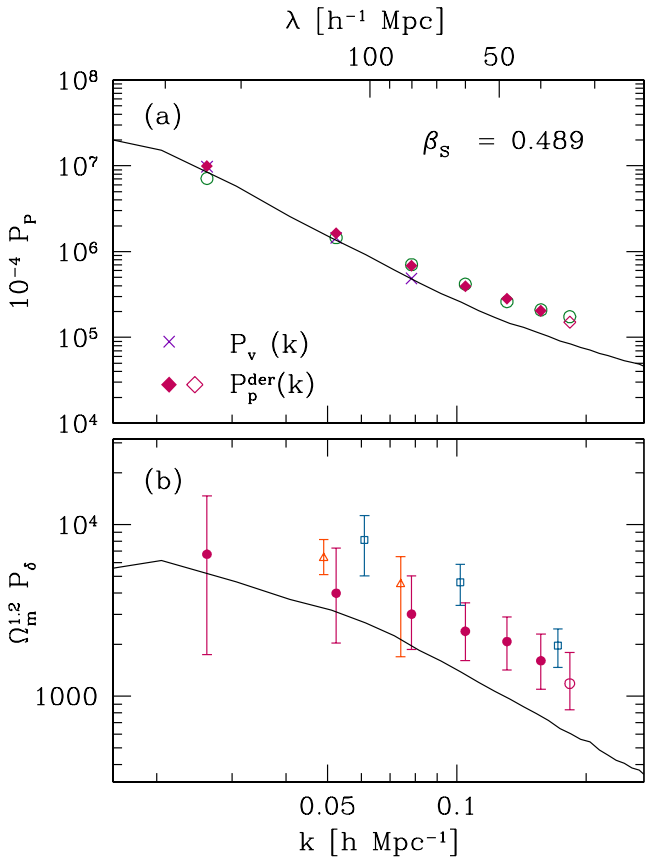,width=88mm,clip=}}
\caption{(Upper) Comparison of the observed momentum PS (open circles)
for the SFI galaxy sample with the momentum PS ($P_p^{\rm der}$; diamonds)
derived from the observed galaxy density PS ($P_{\delta_g}^{\rm obs}$)
for $\beta_{\rm S}=0.489$.
The crosses are the velocity PS given by
$P_v(k)=(H_0\beta_{\rm S})^2 P_{\delta_g}^{\rm obs}(k)/k^2$ for the SFI sample.
(Bottom) Derived matter density PS given by
$\Omega_m^{1.2}P_{\delta}(k)=\beta_{\rm S}^2 P_{\delta_g}^{\rm obs}(k)$
(circles).
For comparisons, the matter density PS measured by \citet{kol97} for the
Mark III sample (open squares) and by P00 for the MAT sample (open triangles)
are plotted. The solid curves are the momentum and matter density PS calculated
from the $N$-body simulation data of the $\Lambda$CDM universe.}
\label{jfig9}
\end{figure}

\begin{deluxetable}{lr}
\tablewidth{0pt}
\tabletypesize{\small}
\tablecaption{Cosmological Parameters Derived from the Momentum
              Power Spectrum Analysis}
\tablecomments{$^\dagger$The rms mass fluctuation of the spiral galaxies
within 8 $h^{-1}$Mpc sphere obtained by integrating the observed SFI density
PS}
\tablecolumns{2}
\tablehead{ Parameters & Estimated Values}
\startdata
 $\beta$ measured from SFI sample
       & $\beta_{\rm S} = 0.49_{-0.05}^{+0.08}$ \\[1mm]
 Matter density
       & $\Omega_m / b_{\rm S}^{5/3} = 0.30_{-0.05}^{+0.09}$ \\[1mm]
 Rms mass fluctuation of spiral galaxies$^\dagger$
       & $\sigma_{8,\rm S} = 1.14_{-0.34}^{+0.33}$ \\[1mm]
 Amplitude of mass fluctuation
       & $\sigma_8 \Omega_m^{0.6} = 0.56_{-0.21}^{+0.27}$ \\[-1mm]
\enddata
\end{deluxetable}

\begin{deluxetable*}{llccc}
%\tabletypesize{\scriptsize}
\tablecolumns{5}
\tablewidth{350pt}
\tablecaption{Comparison of $\beta_{\rm S}$ and $\sigma_8\Omega_m^{0.6}$
              Derived from the Spiral Galaxy Samples}
\tablehead{$\beta_{\rm S}$ & $\sigma_8\Omega_m^{0.6}$ &
           Data & Method & Ref.}
\startdata
   $0.49_{-0.05}^{+0.08}$ & $0.56_{-0.21}^{+0.27}$  
                                    & SFI & Mom. PS & This Study \\[1mm]
   $0.6\pm 0.1$    & \nodata    & SFI, IRAS 1.2 Jy& OME & 1 \\[1mm]
   $0.42\pm0.04$   & \nodata    & SFI, IRAS PSC$z$ & VELMOD & 2 \\[1mm]
   \nodata & $0.82\pm0.12$     & SFI & ML & 3 \\[1mm]
   \nodata & $(0.3\pm0.1)[\Gamma/0.2]^{1/2}$  & SFI & CF & 4 \\[1mm]
   \nodata & $0.63\pm0.08$     & SFI & ML & 5 \\[1mm]
   $0.51_{-0.08}^{+0.13}$    & $0.56\pm0.21$  & MAT & Mom. PS & 6 \\[1mm]
   $0.6\pm0.125$ & \nodata & MAT & ROBUST & 7 \\[1mm]
   $0.50\pm0.10$ & \nodata & Mark III (S), IRAS 1.2 Jy & OME  & 8  \\[1mm]
   $0.49\pm0.07$ & \nodata & Mark III (S), IRAS 1.2 Jy & VELMOD & 9 \\[1mm]
   $0.60_{-0.11}^{+0.13}$ &\nodata & Mark III (S) & $V_{\rm rms}$ & 10 \\[-1mm]
\enddata
\tablerefs{ 
(1) \citet{dac98};
(2) \citet{bra01};
(3) \citet{fre99}; 
(4) \citet{bor00a};
(5) \citet{sil01};
(6) \citet{par00};
(7) \citet{rau00};
(8) \citet{dav96};
(9) \citet{wil97b};
(10) \citet{pad99}} 
\tablecomments{$\Gamma$ is the shape parameter of the CDM models.}
\end{deluxetable*}

Cosmological parameters derived from the momentum PS analysis are summarized
in Table 2, where $\beta_{\rm S}$, $\Omega_m /b_{\rm S}^{5/3}$
($=\beta_{\rm S}^{5/3}$), $\sigma_{8,\rm S}$, and
$\sigma_{8}\Omega_m^{0.6}$ ($=\sigma_{8,\rm S}\beta_{\rm S}$) are listed
with 68\% confidence limits. Here the bias factor for optical spiral galaxies
is defined to be $b_{\rm S}\equiv \sigma_{8,\rm S}/\sigma_{8}$.
Table 3 compares our $\beta_{\rm S}$ and $\sigma_8\Omega_m^{0.6}$ measurements
with the previous results obtained from spiral galaxy samples (SFI, MAT,
and Mark III spirals).
Compared with the previous studies with the SFI data, our $\beta_{\rm S}$
estimate is between those of \citet{bra01} and \citet{dac98}, and
our $\sigma_8\Omega_m^{0.6}$ between the large and the small values of
\citet{fre99} and \citet{bor00a}.
By assuming that the bias factor is a function of $k$, given
by $b_{\rm S}(k)=\left[P_{\delta_g}(k)/P_{\delta}(k)\right]^{1/2}$,
we can infer the PS of mass fluctuation
$\Omega_m^{1.2}P_\delta(k)=\beta_{\rm S}^2(k)P_{\delta_g}(k)$, by multiplying
the observed density PS by $\beta_{\rm S}^2(k)$ at each $k$ (Table 1).
In Figure 9 we plot the PS of mass fluctuation given by
$\Omega_m^{1.2}P_\delta(k)=\beta_{\rm S}^2 P_{\delta_g}^{\rm obs}(k)$ with
$\beta_{\rm S}=0.489$ (filled and open circles in the bottom panel).
For comparisons, the matter density powers measured from the Mark III
(\citealt{kol97}; open squares) and the MAT (P00; open triangles) samples
are also shown.
The MAT mass fluctuation powers are in good agreement with our estimates
while those of Mark III sample are higher than ours on 60--100
$h^{-1}\textrm{Mpc}$ scales.

To assess the effects of MB correction on the $\beta$ estimation,
we have measured the momentum and density PS using the SFI data
that are {\it not} corrected for MB.
The measured density PS has amplitudes very similar to that from the SFI
data with MB correction, but the measured momentum PS has lower amplitudes
at all scales. The $\beta$ parameter, estimated in the same way as described
above, decreases to $0.393$, which is about $2\sigma$ below our estimation
$\beta_{\rm S}=0.489$.
There are systematic effects on $\beta$ estimation from errors in peculiar
velocities and distances of galaxies. To measure the size of systematic
effects, we have calculated $\beta$ parameters using the mock catalogs
with {\it true} distances and peculiar velocities.
The estimated 68\% uncertainty limits are $(-0.13,+0.24)$ and $(-0.048,+0.053)$
for $\beta_{L,{\rm S}}$ and $\beta_{\rm S}$, respectively.
They are very similar to but slightly smaller than those for the realistic
SFI mock catalogs.
This indicates that the uncertainty of $\beta$ for the SFI sample
is still dominated by the statistical effect rather than systematic effects.

\section{Discussion}

We have measured the momentum and density PS from the SFI peculiar velocity
data, and estimated $\beta_{\rm S}=\Omega_m^{0.6}/b_{\rm S}$
parameter and the amplitude of mass fluctuation $\sigma_8\Omega_m^{0.6}$.
Our method is self-consistent because only the SFI peculiar velocity
sample is used without relying on other external velocity
or density fields.
By noting that the momentum PS is accurately related to the galaxy density
PS up to the quasi-linear regime ($\S2$, Fig. 1) and that the density
and momentum PS measured from the same sample fluctuate in a similar way,
we have estimated $\beta$ parameter over a wide range of wavenumber space.
We have determined limits of the linear and the quasi-linear regimes
as $k_L = 0.070$ $h$Mpc$^{-1}$ and $k_{QL} = 0.161$ $h$Mpc$^{-1}$,
and over the ranges $k \lesssim k_L$ and $k \lesssim k_{QL}$
we have estimated $\beta_{L,\rm S}$ and $\beta_{\rm S}(k)$, respectively.
Our $\beta_{\rm S}$ estimation gives stable and consistent values around $0.5$
at scales from 240 $h^{-1}$Mpc to 40 $h^{-1}$Mpc (Fig. 8 and Table 1),
with an average $\beta_{\rm S}=0.49_{-0.05}^{+0.08}$, which translates
into the matter density parameter
$\Omega_m = 0.30_{-0.05}^{+0.09}b_{\rm S}^{5/3}$.
Our measurement is consistent with $\beta$ parameter estimates from other
studies with spiral galaxy samples, and is more accurate.
Especially, the derived parameters are very similar to those from P00
who has applied the momentum PS analysis method to the MAT sample.

\begin{deluxetable*}{cccc}
%\tabletypesize{\scriptsize}
\tablecolumns{4}
\tablewidth{350pt}
\tablecaption{Comparison of $\beta$ and $\sigma_8\Omega_m^{0.6}$ Derived
from the Other Peculiar Velocity Samples}
\tablehead{Parameter & Data & Method & Ref.}
\startdata
\multicolumn{4}{c}{$\beta$}  \\[1mm]
\cline{1-4}
 \\[-1mm]  $0.50\pm0.10$  &  ENEAR, IRAS PSC$z$ & OME ($v$-$v$)& 1  \\[1mm]
   $0.51_{-0.06}^{+0.06}$  & SEcat, IRAS PSC$z$ & UMV ($v$-$v$) & 2  \\[1mm]
   $0.57_{-0.13}^{+0.11}$  & SEcat, IRAS PSC$z$ & UMV ($\delta$-$\delta$) & 2  \\[1mm] 
   $0.50\pm0.04$   & Mark III, IRAS 1.2 Jy & VELMOD ($v$-$v$) & 3  \\[1mm]
   $1.20\pm0.10$   & Mark III, IRAS 1.2 Jy & POTENT ($\delta$-$\delta$) & 4 \\[1mm]
   $0.89\pm0.12$   & Mark III, IRAS 1.2 Jy & POTENT ($\delta$-$\delta$) &  5 \\[1mm]
   $0.50\pm0.06$   & $D_n$--$\sigma$, IRTF, ESO, UGC & $\chi^2$ fit. ($v$-$v$) & 6  \\[1mm]
   $0.39\pm0.17$   & SMAC, IRAS PSC$z$ & $\chi^2$ fit. ($v$-$v$) & 7 \\[1mm]
   $0.40\pm0.15$   & SNIa, IRAS 1.2Jy & $\chi^2$ fit. ($v$-$v$)  & 8 \\[1mm]
   $0.30\pm0.10$   & SNIa, ORS  & $\chi^2$ fit. ($v$-$v$)  & 8 \\[1mm]
   $0.55\pm0.06$   & SNIa, IRAS PSC$z$& $\chi^2$ fit. ($v$-$v$)  & 9 \\[1mm]
   $0.42_{-0.06}^{+0.10}$  &   SBF, IRAS 1.2 Jy & $\chi^2$ fit. ($v$-$v$)  & 10 \\[1mm] 
   $0.26\pm0.08$           &   SBF, ORS  & $\chi^2$ fit. ($v$-$v$)  & 10 \\[1mm]
\cline{1-4}
\\[-1mm] \multicolumn{4}{c}{$\sigma_8 \Omega_m^{0.6}$}  \\[1.2mm]
\cline{1-4}
\\[-1mm]   $0.50_{-0.14}^{+0.25}$ & SCI & $V_{\rm rms}$ & 11 \\[1mm]
   $0.51_{-0.09}^{+0.24}$ for $\Gamma=0.25$ & ENEAR & CF & 12\\[1mm]
   $1.1_{-0.35}^{+0.2}$   & ENEAR  & ML  &  13 \\[1mm]
   $0.88\pm0.15$   & Mark III  & ML  &  14 \\[1mm]
   $0.49\pm0.06$          & Mark III & ML & 15\\[1mm]
   ($0.71$--$0.77$)$\pm0.12$   & Mark III & POTENT & 4 \\[-1mm] 
\enddata
\tablerefs{
(1) \citet{nus01}; 
(2) \citet{zar02b};
(3) \citet{wil98};
(4) \citet{kol97};
(5) \citet{sig98};
(6) \citet{hud94};
(7) \citet{hud04};
(8) \citet{rie97};
(9) \citet{rad04}; 
(10) \citet{bla99};
(11) \citet{bor97}
(12) \citet{bor00b};
(13) \citet{zar01};
(14) \citet{zar97};
(15) \citet{sil01}
}
\end{deluxetable*}

Table 4 summarizes $\beta$ or mass fluctuation measurements
of other studies that use different peculiar velocity samples
(ellipticals, spirals plus ellipticals, and Type Ia supernovae).
In $\beta$ estimation, most of studies depend on the external galaxy
distribution information such as IRAS $1.2$ Jy and PSC$z$ catalogs
for comparing the observed peculiar velocities with that derived from
the galaxy redshift survey data ($v$-$v$ comparison), and give $\beta$
parameters around $0.5$.
However, the POTENT method, one of the $\delta$-$\delta$ comparison methods,
usually gives much higher value of $\beta$ parameter \citep{kol97,sig98}.
Compared with \citet{kol97}, our momentum PS is consistent with their
velocity PS in the linear regime (Fig. 7).
Despite this fact, their estimate of the $\beta$ parameter is much higher
than ours. This is because \citet{kol97} have used the density PS from
other redshift surveys in estimating $\beta$.
Recently, \citet{zar02b} have applied UMV estimator to SEcat, a combination
of the SFI and the ENEAR catalogs, and found consistent $\beta$ estimates
from both $\delta$-$\delta$ and $v$-$v$ comparison methods.
Interestingly, low $\beta$ estimation ($\beta=0.2$--$0.4$) has been reported by
\citet{rie97} and \citet{bla99} who have compared the Optical Redshift
Survey (ORS; \citealt{san95}) with peculiar velocities
from Type Ia supernovae and a surface brightness fluctuation (SBF) survey
of galaxy distances, respectively.

Our measurement of the mass fluctuation level
$\sigma_8\Omega_m^{0.6}=0.56_{-0.21}^{+0.27}$ is larger than that derived
from the Wilkinson Microwave Anisotropy Probe (WMAP; \citealt{ben03})
data analysis. \citet{spe03} have used WMAP data to derive a value of
$\sigma_8\Omega_m^{0.6}=0.44\pm 0.10$.
The reason for the large $\sigma_8\Omega_m^{0.6}$ is that
our measurement of the rms fluctuation of galaxy distribution gives
somewhat large and uncertain value $\sigma_{8,\rm S}=1.14_{-0.34}^{+0.33}$.
By assuming that WMAP cosmological parameters $\sigma_8=0.9$ and
$\Omega_m=0.29$, we obtain
$\sigma_{8,\rm S}=\sigma_8 \Omega_m^{0.6}/\beta_{\rm S}=0.87_{-0.13}^{+0.10}$
for $\beta_{\rm S}=0.489$.
Despite the higher amplitude of mass fluctuation than the WMAP value,
our estimation of the matter density PS $\Omega_m^{1.2}P_{\delta}(k)$
is lower than those of other studies (Table 1 and Fig. 9).
For example, at $k=0.1047~h\textrm{Mpc}^{-1}$ ($\lambda=60~h^{-1}\textrm{Mpc}$),
our PS value is $\Omega_m^{1.2}P_{\delta}(k)=2.51_{-0.94}^{+0.91}\times 10^3$
$(h^{-1}\textrm{Mpc})^3$, which is lower compared with the high-amplitude
power spectra found from the ML analyses of all-sky peculiar velocity
samples such as Mark III, SFI, and ENEAR.
At $k=0.1~h\textrm{Mpc}^{-1}$ scale, the matter density powers
($\Omega_m^{1.2}P_{\delta}$) from the ML analyses are $(4.8\pm1.5)\times 10^3$
(Mark III; \citealt{zar97}), $(4.4\pm 1.7)\times 10^3$ (SFI; \citealt{fre99}),
and $(6.5\pm3)\times 10^3$ $(h^{-1}\textrm{Mpc})^3$ (ENEAR; \citealt{zar01}).
Those analyses also give the high mass fluctuation levels
($\sigma_8\Omega_m^{0.6}$) around 0.8--1.1 (Table 3 and 4).
On the contrary, \citet{sil01} found a power deficiency at
$k=0.1~h\textrm{Mpc}^{-1}$ from the ML analysis of the SFI sample, and
their density PS and $\sigma_8\Omega_m^{0.6}$ are consistent with our results.

As shown in Figure 8, the $\beta$ in the quasi-linear regime is more accurate
than that in the linear-regime.
This implies that if we rely on the linear regime where the density and
velocity PS are simply related by the linear gravitational instability theory
as in equation (5), we need peculiar velocity data with large survey volume
for accurate determination of the $\beta$ parameter.
On the other hand, if we use the full information of the momentum field
from linear to quasi-linear regimes where the relation
between the density and momentum PS are accurately known,
a peculiar velocity sample gives more accurate $\beta$ compared with
that estimated only in the linear regime.
For the best $\beta$ determination, it is essential to derive a more accurate
relation that connects the density PS with the momentum PS beyond the
limit of the quasi-linear regime by applying the higher order perturbation
theory.

\begin{acknowledgements}
We gratefully acknowledge to Dr. Riccardo Giovanelli and Dr. Martha Haynes
for providing the SFI galaxy sample.
CGP acknowledges valuable comments from Kin-Wang Ng and Juhan Kim.
This work was supported by the Astrophysical Research Center for the Structure
and Evolution of the Cosmos (ARCSEC) of the Korea Science and Engineering
Foundation (KOSEF) through Science Research Center (SRC) program.
\end{acknowledgements}


\begin{thebibliography}{}
\bibitem[Bennett et al.(2003)]{ben03} Bennett, C.L., et al. 2003, \apjs, 148, 1
\bibitem[Bernardeau et al.(2002)]{ber02} Bernardeau, F., Colombi, S.,
   Gazta\~naga, E., \& Scoccimarro, R. 2002, Phys. Rep. 367, 1
\bibitem[Bertschinger et al.(1990)]{ber90} Bertschinger, E., Dekel, A., Faber,
   S.M., Dressler, A., \& Burstein, D. 1990, \apj, 364, 370
\bibitem[Blakeslee et al.(1999)]{bla99} Blakeslee, J.P., Davis, M.,
   Tonry, J.L., Dressler, A., \& Ajhar, E.A. 1999, \apj, 527, L73
\bibitem[Borgani et al.(1997)]{bor97} Borgani, S., da Costa, L.N., Freudling,
   W., Giovanelli, R., Haynes, M.P., Salzer, J., \& Wegner, G. 1997, \apj,
   482, L121
\bibitem[Borgani et al.(2000a)]{bor00a} Borgani, S., da Costa, L.N., Zehavi, I.,
   Giovanelli, R., Haynes, M.P., Freudling, W., Wegner, G., \& Salzer, J.J.
   2000a, \apj, 119, 102
\bibitem[Borgani et al.(2000b)]{bor00b} Borgani, S., Bernardi, M., da Costa,
   L.N., Wegner, G., Alonso, M.V., Willmer, C.N.A., Pellegrini, P.S., \&
   Maia, M.A.G. 2000b, \apj, 537, L1
\bibitem[Branchini et al.(2001)]{bra01} Branchini, E., et al. 2001, \mnras,
   326, 1191
\bibitem[Branchini et al.(1999)]{bra99} Branchini, E., et al. 1999, \mnras,
   308, 1
\bibitem[da Costa et al.(2000a)]{dac00a} da Costa, L.N., Bernardi, M., Alonso,
   M.V., Wegner, G., Willmer, C.N.A., Pellegrini, P.S., Maia, M.A.G., \&
   Zaroubi, S. 2000a, \apj, 537, L81
\bibitem[da Costa et al.(2000b)]{dac00b} da Costa, L.N., Bernardi, M., Alonso,
   M.V., Wegner, G., Willmer, C.N.A., Pellegrini, P.S., Rit\'e, C., \&
   Maia, M.A.G. 2000b, \aj, 120, 95
\bibitem[da Costa et al.(1996)]{dac96} da Costa, L.N., Freudling, W., Wegner,
   G., Giovanelli, R., Haynes, M.P., \& Salzer, J.J. 1996, \apj, 468, L5
\bibitem[da Costa et al.(1998)]{dac98} da Costa, L.N., Nusser, A., Freudling,
   W., Giovanelli, R., Haynes, M.P., Salzer, J.J., \& Wegner, G. 1998, \mnras,
   299, 425
\bibitem[Davis et al.(1996)]{dav96} Davis, M., Nusser A., \& Willick, J.A.
   1996, \apj, 473, 22
\bibitem[Dekel et al.(1999)]{dek99} Dekel, A., Eldar, A., Kolatt, T., Yahil, A.,
   Willick, J.A., Faber, S.M., Courteau, S., \& Burstein, D. 1999, \apj, 522, 1
\bibitem[Eisenstein \& Hu(1999)]{eis99} Eisenstein, D.J., \& Hu, W. 1999,
   \apj, 511, 5
\bibitem[Feldman et al.(2003)]{fel03} Feldman, H., et al. 2003, \apj, 596, L131
\bibitem[Freudling et al.(1995)]{fre95} Freudling, W., da Costa, L.N.,
   Wegner, G., Giovanelli, R., Haynes, M.P., \& Salzer, J.J. 1995, \aj, 110, 920
\bibitem[Freudling et al.(1999)]{fre99} Freudling, W., et al. 1999, \apj,
   523, 1
\bibitem[Giovanelli et al.(1994)]{gio94} Giovanelli, R., Haynes, M.P., 
   Salzer, J.J., Wegner, G., da Costa, L.N., \& Freudling, W. 1994, \aj,
   107, 2036 
\bibitem[Giovanelli et al.(1997a)]{gio97a} Giovanelli, R., Haynes, M.P., 
   Herter, T., Vogt, N.P., da Costa, L.N., Freudling, W., Salzer, J.J., \&
    Wegner, G. 1997a, \aj, 113, 53
\bibitem[Giovanelli et al.(1997b)]{gio97b} Giovanelli, R., Haynes, M.P.,
   Herter, T., Vogt, N.P., Wegner, G., Salzer, J.J., da Costa, L.N., \&
   Freudling W. 1997b, \aj, 113, 22
\bibitem[G\'orski et al.(1989)]{gor89} G\'orski, K., Davis, M., Strauss, M.A.,
   White, S.D.M., \& Yahil, A. 1989, \apj, 344, 1
\bibitem[Haynes et al.(1999a)]{hay99a} Haynes, M.P., Giovanelli, R., 
   Chamaraux, P., da Costa, L.N., Freudling, W., Salzer, J.J., \&  Wegner, G.
   1999a, \aj, 117, 2039
\bibitem[Haynes et al.(1999b)]{hay99b} Haynes, M.P., Giovanelli, R., Salzer, 
   J.J., Wegner, G., Freudling, W., da Costa, L.N., Herter, T., \& Vogt, N.P.
   1999b, \aj, 117, 1668
\bibitem[Hockney \& Eastwood(1981)]{hoc81} Hockney, R.W., \& Eastwood, J.W.
   1981, Computer Simulation Using Particles (New York: McGraw-Hill, 1981)
\bibitem[Hudson(1994)]{hud94} Hudson, M.J. 1994, \mnras, 266, 475
\bibitem[Hudson et al.(2004)]{hud04} Hudson, M.J., Smith, R.J., Lucey, J.R.,
   \& Branchini, E. 2004, \mnras, 352, 61
\bibitem[Kolatt \& Dekel(1997)]{kol97} Kolatt, T., \& Dekel, A. 1997, \apj,
   479, 592
\bibitem[Lynden-Bell et al.(1988)]{lyn88} Lynden-Bell, D., Faber, S.M.,
   Burstein, D., Davies, R.L., Dressler, A., Terlevich, R.J., \& Wegner, G.
   1988, \apj, 326, 19
\bibitem[Mathewson et al.(1992)]{mat92} Mathewson, D.S., Ford, V.L., \&
   Buchhorn, M. 1992, \apjs, 81, 413 (MAT)
\bibitem[Nusser et al.(2001)]{nus01} Nusser, A., da Costa, L.N., Branchini, E.,
   Bernardi, M., Alonso, M.V., Wegner, G., Willmer, C.N.A., \& Pellegrini, P.S.
   2001, \mnras, 320, L21
\bibitem[Nusser \& Davis(1995)]{nus95} Nusser, A., \& Davis, M., 1995, \mnras,
   276, 1391
\bibitem[Padilla \& Lambas(1999)]{pad99} Padilla, N., \& Lambas, D.G. 1999,
   \mnras, 310, 21
\bibitem[Park(1990)]{par90} Park, C. 1990, \mnras, 242, 59p
\bibitem[Park(1997)]{par97} Park, C. 1997, J. Korean Astron. Soc., 30, 191
\bibitem[Park(2000)]{par00} Park, C. 2000, \mnras, 319, 573 (P00)
\bibitem[Park et al.(1992)]{par92} Park, C., Gott, J.R., \& da Costa, L.N. 1992,
   \apj, 392, L51
\bibitem[Park et al.(1994)]{par94} Park, C., Vogeley, M.S., Geller, M.J.,
   \& Huchra, J.P. 1994, \apj, 431, 569 
\bibitem[Peebles(1980)]{pee80} Peebles, P.J.E. 1980, The Large-Scale Structure
   of the Universe (Princeton University Press)
\bibitem[Peebles(1993)]{pee93} Peebles, P.J.E. 1993, Principles of Physical
   Cosmology (Princeton University Press)
\bibitem[Radburn-Smith et al.(2004)]{rad04} Radburn-Smith, D.J., Lucey, J.R.,
   \& Hudson, M.J. 2004, \mnras, 355, 1378
\bibitem[Rauzy \& Hendry(2000)]{rau00} Rauzy, S., \& Hendry, M.A. 2000,
   \mnras, 316, 621
\bibitem[Riess et al.(1997)]{rie97} Riess, A.G., Davis, M., Baker, J., \&
   Kirshner, R.P. 1997, \apj, 488, L1
\bibitem[Santiago et al.(1995)]{san95} Santiago, B.X., Strauss, M.A., Lahav,
   O., Davis, M., Dressler, A., \& Huchra, J.P. 1995, \apj, 446, 457
\bibitem[Saunders et al.(2000)]{sau00} Saunders, W., et al. 2000, \mnras,
   317, 55
\bibitem[Schmidt(1968)]{sch68} Schmidt, M. 1968, \apj, 151, 393
\bibitem[Sigad et al.(1998)]{sig98} Sigad, Y., Eldar, A., Dekel, A., Strauss,
   M.A., \& Yahil, A. 1998, \apj, 495, 516
\bibitem[Silberman et al.(2001)]{sil01} Silberman, L., Dekel, A., Eldar, A.,
   \& Zehavi, I. 2001, \apj, 557, 102
\bibitem[Spergel et al.(2003)]{spe03} Spergel, D.N., et al. 2003, \apjs, 148,
   175
\bibitem[Strauss \& Willick(1995)]{str95} Strauss, M.A., \& Willick, J.A. 1995,
   Phys. Rep., 261, 271
\bibitem[Tegmark et al.(2004)]{teg04} Tegmark, M., et al. 2004, \apj, 660, 702
\bibitem[Vogeley et al.(1992)]{vog92} Vogeley, M.S., Park, C., Geller, M.J.,
   \& Huchra, J.P. 1992, \apj, 391, L5
\bibitem[Watkins et al.(2002)]{wat02} Watkins, R., Feldman, H.A.,
   Chambers, S.W., Gorman, P., \& Melott, A.L. 2002, \apj, 564, 534
\bibitem[Willick et al.(1997a)]{wil97a} Willick, J.A., Courteau, S.,
   Faber, S.M., Burstein, D., Dekel, A., \& Strauss, M.A. 1997, \apjs, 109, 333
\bibitem[Willick \& Strauss(1998)]{wil98} Willick, J.A., \& Strauss, M.A.
   1998, \apj, 507, 64
\bibitem[Willick et al.(1997b)]{wil97b} Willick, J.A., Strauss, M.A., Dekel, A.,
   \& Kolatt, T. 1997, \apj, 486, 629
\bibitem[Zaroubi(2002)]{zar02a} Zaroubi, S. 2002, in
   Proceedings of XIII Rencontres de Blois, Frontiers of the Universe,
   ed. L.M. Celnikier et al. p.65 (astro-ph/0206052)
\bibitem[Zaroubi et al.(2001)]{zar01} Zaroubi, S., Bernardi, M., da Costa, L.N.,
   Hoffman, Y., Alonso, M.V., Wegner, G., Willmer, C.N.A., \& Pellegrini, P.S.
   2001, \mnras, 326, 375
\bibitem[Zaroubi et al.(2002)]{zar02b} Zaroubi, S., Branchini, E., Hoffman, Y.,
   \& da Costa, L.N. 2002, \mnras, 336, 1234
\bibitem[Zaroubi, Hoffman, \& Dekel(1999)]{zar99} Zaroubi, S., Hoffman, Y.,
   \& Dekel, A. 1999, \apj, 520, 413
\bibitem[Zaroubi et al.(1997)]{zar97} Zaroubi, S., Zehavi, I., Dekel, A.,
   Hoffman, Y., \& Kolatt, T. 1997, \apj, 486, 21
\end{thebibliography}
\end{document}